\documentclass[11pt]{article}
\usepackage{graphicx}
\begin{document}
%

\begin{center}
{\large \bf The Standard Model}

\vskip.5cm

W-Y. Pauchy Hwang\footnote{Correspondence Author;
 Email: wyhwang@phys.ntu.edu.tw; A completed version
 (Version 3) of arXiv:1304.4705v1 [hep-ph] 17 April
 2013 (the original version).}
 \\
{\em Asia Pacific Organization on Cosmology and Particle Astrophysics, \\
Institute of Astrophysics, Center for Theoretical Sciences,\\
and Department of Physics, National Taiwan University,
     Taipei 106, Taiwan}
\vskip.2cm


{\small(April 17, 2013; updated on August 25, 2015; August 25, 2017)}
\end{center}

\begin{abstract}
We declare that we live in the quantum 4-dimensional
Minkowski space-time with, {\it via the gauge
principle,} the force-fields gauge-group structure,
$SU_c(3) \times SU_L(2) \times U(1)$ applied to the
quark world or $SU_L(2) \times U(1) \times
SU_f(3)$ applied to the lepton world, built-in
from the very beginning. The quark world with
the triplets from the "quark" group $SU_Q(3)$ (used to
be called "flavor $SU(3)$ symmetry"), which is of
nuclear sizes and is protected by $SU_c(3) \times
SU_L(2) \times U(1)$ (i.e. the (123) symmetry), can
be seen by our world, while the lepton world, as of
atomic sizes and protected by $SU_L(2) \times U(1)
\times SU_f(3)$ or another (123) symmetry, can also
be seen, as singlets of the quark group $SU_Q(3)$,
by our world. Apart from the "ignition" term, the
entire Standard Model is dimensionless and massless
in the quantum 4-dimensional Minkowski space-time;
that is, all couplings are dimensionless and there
are no mass terms in the quantum 4-dimensional
Minkowski space-time.

Therefore, there exist the well-studied $3^\circ \,K$
cosmic microwave background (CMB) and the now clustered
neutrino halos formed from cosmic background (CB)
neutrinos (of three flavors, and antineutrinos), both
present in the overall background of our Universe.
In theory, it yields neutrino oscillations, as some
lepton-flavor-violating processes, in a natural manner.

\bigskip

{\parindent=0pt PACS Indices: 12.60.-i (Models beyond the standard
model); 98.80.Bp (Origin and formation of the Universe); 12.10.-g
(Unified field theories and models).}
\end{abstract}

\medskip

\section{The Excerpts}

We declare that our world is the quantum 4-dimensional
Minkowski space-time with, via the gauge principle, the
force-fields gauge-group structure, $SU_c(3) \times
SU_L(2) \times U(1)$ at the fermi scale or $SU_L(2)
\times U(1) \times SU_f(3)$ at the anstron scale,
implemented at the very beginning.

The {\it matter} of our world has triplets of the quark
group $SU_Q(3)$ and singlets of the quark group
$SU_Q(3)$, each in terms of one left-handed $SU_L(2)$
doublet and two right-handed $SU_L(2)$ singlets,
while the triplets (quarks) of the quark group have the
gauge symmetry $SU_c(3) \times SU_L(2) \times U(1)$
and the singlets (leptons) of the quark group have the
gauge symmetry $SU_L(2) \times U(1) \times SU_f(3)$.
The {\it overall background} is the quantum 4-dimensional
Minkowski space-time with, via the gauge principle, the
force-field gauge-group structure suitably built-in at
the very beginning.

The (123) gauge symmetry, under $SU_c(3) \times SU_L(2)
\times U(1)$, and the other (123) gauge symmetry, under
$SU_L(2) \times U(1) \times SU_f(3)$, are explained
later on in this paper - c.f., Section 7 when the
meanings of the notations become clear.

{\it Mathematically, the Standard Model \cite{definition}
is a group theory - a group of the elements, triplets and
singlets, of the quark group $SU_Q(3)$; they interact
among themselves through the gauge symmetry, either
$SU_c(3) \times SU_L(2) \times U(1)$ at the fermi sizes
or $SU_L(2) \times U(1) \times SU_f(3)$ at the atomic
sizes. They interact on the background of the quantum
4-dimensional Minkowski space-time.}

\medskip

\section{Introduction}

According to Newton's doctrine, the motion of an
object has to be specified by both its space-time
point (i.e., coordinates) and its behavior at a
nearby point (thus, its velocities). The force
information, which controls the change of the
velocity, is characterized by the gauge
principle, $\partial_\mu \to D_\mu$. In modern
days, the difference, $D_\mu-\partial_\mu$, is
expressed by the gauge fields. Thus, we still
stick to Newton's doctrine and the various
gauge fields should be characterized at the
very beginning of the game.

The next thing in the game is how to introduce
the matter worlds, i.e., the quark world (at the
fermi scale) and the lepton world (at the atomic
scale). The gauge principle is expected
to act on some basic unit of matter - these units
of matter are expressed in terms of the entries
from the quark world, or, the entries from the
lepton world. As seen below, the three
entries, two right-handed units of matter
and one left-handed unit of matter, together
are used to define the lepton world; likewise,
the three similar entries, in terms of the
triplets of some symmetry group $SU_Q(3)$ (to
be explained later on), are used to define
the quark world.

We believe that the language will be used, for
centuries to come, to describe the smallest units
of matter, including electrons, neutrinos, and
quarks, on the basis of Einstein's relativity
principle and the quantum principle.

Nowadays it is a general belief (of others so far, but of not the
view of this paper) that there are three generations of quarks
and three generations of leptons, at the level of the so-called
"point-like Dirac particles". According to another newly
established belief in Cosmology, the content of the current
Universe would be 25\% in the dark matter while only 5\% in the
visual ordinary matter, the latter described by the "minimal
Standard Model" (mSM). In this language, the dark-matter particles
are supposed to be described by some real Standard Model. Indeed,
there is certain urgent need for a truly Standard Model, which
also accommodates those phenomena which are currently classified
as "beyond the Standard Model".

We could call the above statements as "the old standard wisdom"
(i.e., before the year of 2017). They would appear in talks,
in papers, etc., as of today (2017). The Standard Model, the
name that was invented by S. Weinberg in 1970's, of this paper
deviates from the above standard wisdom in that there is new
family gauge group $SU_f(3)$ on top of the well-known
force-fields gauge group $SU_c(3) \times SU_L(2) \times U(1)$.
The three "generations" of leptons are easily explained away
and neutrino oscillations among three "generations" can be
easily understood. In the quark world, all the entries are
the triplets of the quark group $SU_Q(3)$, such that the old
notion of "generations" also never appear.

What we are doing in this paper is to describe the
Standard Model that is based on Einstein's relativity
principle and the quantum principle (established
in the 20th Century) and that describes the {\it
smallest units of matter}, including electrons,
neutrinos, and quarks. The Standard Model
\cite{definition} is, apart from the SSB "ignition"
term, both the originally massless theory and the
dimensionless theory. Both the gauge principle and
the principle of renormalizability are now elevated
to the status of the (mathematical) principles.
(SSB means "spontaneous symmetry breaking".)

Nowadays we all know that we have firmly established the
validity of Einstein's relativity principle and the
quantum principle, the so-called two foundation pillars
established as of the 20th Century. Starting from the
beginning of the 21st Century, we should take it very
seriously that there {\it exist} the {\it smallest}
units of matter, such as the electrons, neutrinos,
quarks, etc., a finite number of them. The behaviors
of these smallest units of matter are described by
the Standard Model, which we try to discuss in
the present paper.

We insist on the quantum principle, in the sense that
electrons, neutrinos, and quarks satisfy the Dirac
equations, thereby satisfying the mysterious closed
anti-commuting Dirac algebra, and they are fermions
satisfying Pauli's exclusion principle when stacking
up. Thus, these are truly the quantum phenomena,
although this used to be a rather murky area as far
as the quantum principle is concerned.

Theoretically, we suggest that we live in the quantum 4-dimensional
Minkowski space-time with, {\it via the gauge principle,} the
force-fields gauge-group structure, $SU_c(3) \times SU_L(2) \times
U(1)$ applied in the triple quark world and $SU_L(2) \times U(1)
\times SU_f(3)$ applied in the lepton world, built-in from
the very beginning. In this overall background, the triple quark
world is observed, of nuclear sizes, because of the (123) symmetry
(i.e., under $SU_c(3) \times SU_L(2) \times U(1)$), while the
lepton world is observed, of atomic sizes, in view of the other
(123) symmetry (i.e., under $SU_L(2) \times U(1) \times SU_f(3)$).
This gives us "our world" or "our Universe" \cite{definition}.

The gauge principle helps to link the forces with
our Minkowski space-time, since it tells us how to move
the coordinates $x_\mu$ to a nearby point $x_\mu +
\delta x_\mu$. The substitution, $\partial_\mu \to D_\mu$
for each basic unit of matter, reflects the information
of the gauge principle.

Here $SU_f(3)$ stands for the $SU_{family}(3)$ family gauge theory,
rather than $SU_{flavor}(3)$ (to be named later as $SU_Q(3)$, to
avoid any further confusion) which derives from the isospin
symmetry - the latter {\it not} a gauge theory. Presumably
the size consideration forces the choice of strong $SU_c(3)$
over family $SU_f(3)$ for the triple quark world while, for the
lepton world, we don't understand why the strong $SU_c(3)$ is
completely shut off. In the lepton world, we insist on the
symmetry $SU_L(2) \times U(1) \times SU_f(3)$, for the sake of
the well-behaved limits at the small distances.

Among the detailed efforts to include the $SU_f(3)$ gauge group,
we should mention those of T. Yanagida \cite{Yanagida} and of
Yue-Liang Wu \cite{YLWu}. In our language, the force-fields background,
the triple quark world, and the lepton world are separate entities
to begin with - they have their own individual ranges and other
characteristics. Each of them are well-behaved, mathematically
and physically. Thus, it is more natural that $SU_f(3)$ covers
only the lepton world - making it free of Landau ghosts and
making it asymptotically free. Thus, our $SU_f(3)$ does not
cover the triple quark world, unlike \cite{Yanagida, YLWu}.
The applicability of $SU_f(3)$ to leptons, or quarks, or both,
is the issue. In fact, the gauge principle forces us to
make one choice, and only one choice, from either of
$SU_c(3)$ and $SU_f(3)$.

There are various reasons why the $SU_f(3)$ should be with
the lepton world. Without the $SU_f(3)$, Landau's ghosts
would kill it as a consistent theory, as conceptually it would
blow up as the distance goes to zero. With the $SU_f(3)$
in the lepton world, the left-handed
leptons have to put together to form a $(3,2)$ multiplet; that
would call for a $(3,2)$ family Higgs multiplet; and then
to make every $SU_f(3)$ gauge boson massive we require a
$(3,1)$ purely family Higgs multiplet. It turns out that
it is harmless to have the charged partner in $\Phi(3,2)$
- they would not go through the spontaneous symmetry
breaking (SSB). In terms of the degrees of freedom (DOF),
it turns out to be just right. It accommodates three
kinds of neutrinos to oscillate among themselves.

Besides all these, the complex scalar field $\phi(x)$
alone cannot exist owing to the self-repulsive
interaction $\lambda (\phi^\dagger \phi)^2$. So,
we need $\Phi(3,2)$ and $\Phi(3,1)$ to guarantee
the existence of the Standard-Model (SM) Higgs.
It is indeed amazing.

\medskip

\section{The Entries and the Background of the Standard Model}

In Newton's doctrine, it is {\it not} sufficient
to specify an object by just giving the space-time
point $x_\mu$ - instead, a complete description
also  needs the velocity information in terms of
$\partial_\mu$. Thus, the lagrangian is a
functional of the coordinates and their first
derivatives (velocities). The gauge principle
means to replace the first derivatives by the
gauge-invariant derivatives $D_\mu$, with
the differences $D_\mu-\partial_\mu$
characterized linearly by the gauge
fields (potentials).

Thus, the background of the Standard Model contains
not only the space-time coordinates but also their
changes, i.e., the velocities; the changes in
velocities through the gauge principle which contains
the gauge-fields force information are required. So,
this defines "the background" or "the overall
background".

We have two kinds of the inputs (or the entry
points): One input is at the fermi or $10^{13}
\,cm$ sizes while the other input is at the
atomic or $10^{-8}\, cm$ sizes.

The lepton world, or the atomic world,
{\it via the gauge principle,}
sees the force fields characterized by
the gauge group $SU_L(2) \times U(1)
\times SU_f(3)$. The typical sizes
are of $(10^{-8}\,cm)^3$.

The quark world with the entries
from members of the "quark" group $SU_Q(3)$
(used to be called "flavor $SU(3)$
symmetry"), or the nuclear world, {\it
via the gauge principle,} sees the force
fields characterized by the gauge group
$SU_c(3) \times SU_L(2) \times U(1)$.
The typical sizes are of $(10^{-13}\,cm)^3$.

To make it clear, the lepton world is the singlet
members of the quark group $SU_Q(3)$. One should
reckon the supreme status of the group $SU_Q(3)$
- it should cover everything. By our Universe,
the Standard Model \cite{definition} should cover
everything, including its mathematical meaning as
a group. We call $SU_Q(3)$ the "quark" group mainly
because the entry points of the quark world are
the triplets of the group $SU_Q(3)$.

To be precise, the lepton world is defined by one
left-handed $SU_Q(3)$ singlet and two right-handed
$SU_Q(3)$ singlets. Meanwhile, the quark world is
defined by one left-handed $SU_Q(3)$ triplet and
two right-handed $SU_Q(3)$ triplets. These are
the inputs of the Standard Model - we are trying
to use the group theory to say things.

In the mathematical language, it may be
appropriate to introduce the concept of
the quark group $SU_Q(3)$ - a generalization
of the flavor $SU(3)$ symmetry. It should
apply to the lepton world, as well. That
makes the group concept omnipotent and
the members of the group have some certain
common characteristics to begin with. Mathematics
and physics have some common traits, after all.

We wish to remark that the meaning of the
"generations" disappears altogether in this
Standard Model - in the quark world the triplet
members of the quark group $SU_Q(3)$ serve as
the basic entries, while in the lepton world
the "generations" are replaced by the
fermion members of the family gauge group
$SU_f(3)$. Thus, "the three generations"
are replaced by the entries of the gauge
group $SU_f(3)$ while "the three generations"
are replaced by the triplet entries of the
quark group $SU_Q(3)$ - it turns out that
we no longer need to talk about the
"generations".

Both the matter worlds, the lepton world at the
atomic scale and the quark world at the fermi
scale, are asymptotically free, and thus free
of Landau ghosts. In fact, the overall consistency
of the Standard Model \cite{definition} guarantees
its mathematical existence. In other world, we can
analyze its (mathematical) existence by studying
its overall consistency.

In view of the fundamental importance, we would like
to suggest to promote "the gauge principle" and "the
principle of renormalizability" to the full status of
the principles, its meaning more as the mathematical
principles (than in the physics principles).

\medskip

\section{The Principle of Renormalizability}

A real strange thing happens in the 4-dimensional Minkowski
space-time \cite{definition} - a story which we should have
known but so far didn't. A complex scalar field $\phi(x)$
should have the self-interaction $\lambda (\phi^\dagger
\phi)^2$ with a dimensionless (positive) $\lambda$ cannot
be observed if being alone - since a positive $\lambda$
means self-repulsive. It is dimensionless so it is determined
by the space-time, but {\it not} by the field itself.
Maybe $\lambda={1\over 8}$ but we can't prove it so far.
Thus, the "related" complex scalar fields $\Phi(1,2)$
(Standard-Model Higgs), $\Phi(3,2)$ (mixed family Higgs),
and $\Phi(3,1)$ (purely family Higgs) come together to
sing a chorus - an SSB chorus with the various
force-fields gauge fields.

The self-repulsive term $\lambda (\phi^\dagger \phi)^2$
($\lambda > 0$) is completely acceptable in terms
of renormalizability, but it precludes the existence
of the field $\phi(x)$. Thus, we need the existence of
another "related" complex scalar field $\phi_2(x)$ such
that a mutual-attractive term $-4 \lambda
(\phi_1^\dagger \phi_2) \cdot (\phi_2^\dagger \phi_1)$
exist to save the situation.

The logic is rather simple. We're living in the
quantum 4-dimensional Minkowski space-time.
Through the gauge principle, we (i.e., in terms
of the basic units of matter) are also connected
with the various force fields (gauge fields). We
have complex scalar fields, to make up the gauge
fields (to make them massive). The so-called
"matter", in terms of Dirac fields, including
electrons, neutrinos, and quarks, all exist
there. The complex scalar fields have the
similar characteristics as the overall
background - the quantum 4-dimensional
Minkowski space-time. They are self-repulsive
and cannot exist alone by itself. But they
are "related" - the SM Higgs $\Phi(1,2)$,
the mixed family Higgs $\Phi(3,2)$, and the
purely family Higgs $\Phi(3,1)$, "related"
to lower their total energy and to make
the SM and family gauge bosons massive.
This makes up the so-called "overall
background". The triple quark world (with
the entries as members of the quark group
$SU_Q(3)$), with the (123) gauge symmetry,
is accepted by this background. Further,
the lepton world, with another (123) gauge
symmetry and with a much bigger scale
(sizes), is also accepted by the same
background. We are talking about the origin
of fields or of point-like particles, talking
about the "renormalizable" Standard Model
\cite{Fields, definition}. We should figure
out a way to think about our entire world,
even if we might not be on the right track.

At the end, the Standard Model is a consistent
language of mathematics - each field represents
a certain particle, no more redundant field
(particle). There is the "overall" background,
namely, the quantum 4-dimensional Minkowski
space-time (a physical system) with, {\it via
the gauge principle,} the force-fields
gauge-group structure suitably built-in from the
very beginning, plus the triple quark world and the
lepton world.

\medskip

\section{The Gauge Principle}

The {\it "basic units of matter"}, which are based on the
right-handed Dirac components or left-handed Dirac components
but have their group assignments {\it explicit} under the
group $SU_c(3) \times SU_L(2) \times U(1) \times SU_f(3)$,
would be more appropriate than the so-called "building blocks
of matter". Moreover, each basic unit derives from one
kinetic-energy term, $-{\bar \Psi}_R \gamma_\mu \partial_\mu \Psi_R$
or $-{\bar\Psi}_L\gamma_\mu \partial_\mu \Psi_L$, and from only one
such term. The gauge principle tells us that $\partial_\mu$ is to be
replaced by some "gauge-invariant" derivative $D_\mu$. It is clear that
this would be an economical way to write down a theory in the globally
consistent manner.

In fact, the gauge group would be $SU_c(3) \times SU_L(2) \times
U(1)$ or $SU_L(2) \times U(1) \times SU_f(3)$, since the two
$SU(3)$, if showing up simultaneously, would contradict each
other. Moreover, if we introduce the "quark" group $SU_Q(3)$,
used to be called "the flavor $SU(3)$ symmetry", the entries
in the theory should be the triple $(d'_R, s'_R, b'_R)$, etc.,
instead of $d'_R$ individually, and so on - thus, the three
"generations" in the quark case is a misnomer.

These {\it basic} units of matter turn out to be the
{\it smallest} units of matter - so, we may use
these names interchangeably. For example, an electron,
being one basic unit of matter, is indivisible by
any means. So are the other basic units of matter,
such as neutrinos, quarks, etc.

The most important aspect is that the kinetic energy term
$-{\bar \Psi} (x) \gamma_\mu \partial_\mu \Psi(x)$
can appear {\it once and only once} in the lagrangian
- this fact was given in Newton's classic doctrine in
the description of particle's motions. The gauge
principle $\partial_\mu \to D_\mu$ then implies
that there is {\it one and only one} $D_\mu$. The
difference, $D_\mu-\partial_\mu$, is a measure
of the influence of the force(s). This is why, in
statement of the Standard Model, the force-fields
gauge-group structure has to be built in from the
very beginning - it has to precede the existence
of the quark world or of the lepton world.

Clearly, we may split the ordinary-matter world into
the "quark" world and the "lepton" world, as the ranges of their
"livings" are quite different from each other - the quark world
is at the fermi scale while the lepton world at the atomic
scale. At low enough temperature, the systems of quarks
(and anti-quarks) are confined while systems in leptons could
roam anywhere.

We know that the right-handed neutrinos do not appear in the minimal
Standard Model - so, $(\nu_{\tau R},\, \nu_{\mu R},\, \nu_{e R})$
would make a perfect triplet under $SU_f(3)$.
In a recent paper \cite{HwangYan}, we proposed to put $((\nu_\tau,\,
\tau)_L,\,(\nu_\mu,\,\mu)_L, \,(\nu_e,\,e)_L)$ $(columns)$ ($\equiv
\Psi_L(3,2)$) as the $SU_f(3)$ triplet and $SU_L(2)$ doublet. The next
question is how to assign right-handed $\tau_R$, $\mu_R$, and $e_R$
under $SU_f(3)$. We could write down the mass term for the charged
leptons - if they are singlets or $(\tau_R,\, \mu_R, \, e_R)$ is
an $SU_f(3)$ triplet? The singlet assignment can be ruled out
since there are undesirable crossed terms in, e.g.,
$\Psi_L(\bar 3,2) e_R \Phi(3,2)$. So, three of them would
better form a triplet - $\Psi_R^C(3,1)$. In fact, This
completes the list of "the basic units" in the lepton world.

{\it The lepton world is constructed from three
group elements $(\nu_{\tau R}, \nu_{\mu R}, \nu_{e R})$,
$(\tau_R, \mu_R, e_R)$, and $((\nu_\tau, \tau)_L,
\,(\nu_\mu,\mu)_L,\,(\nu_e,e)_L)$ (columns). The
group is $SU_L(2) \times U(1) \times SU_f(3)$.
This should be regarded as the starting point of
the lepton world.}

Since we wish to propose the $SU_f(3)$ family gauge theory
as a way to understand why there are three generations, it requires all
additional particles, i.e., (eight) gauge bosons and (four) residual family
Higgs, very massive. In the proposal of Hwang and Yan \cite{HwangYan}, we
would have one Standard-Model Higgs field $\Phi(1,2)$, one complex family
Higgs triplet $\Phi(3,1)$, and another family triplet-doublet complex scalar
fields $\Phi(3,2)$. Amazingly enough, two neutral complex triplets,
$\Phi(3,1)$ and $\Phi^0(3,2)$ would indeed undergo the desired Higgs mechanism
and the three charged scalar fields would remain massive, i.e. there is no
spontaneous symmetry breaking (SSB) in the charged Higgs sector.

So far, we have decided on the basic units of matter - those
left-handed and right-handed objects (quarks or leptons);
the gauge group is chosen to be $SU_c(3) \times SU_L(2)
\times U(1) \times SU_f(3)$. As we said, we live in the
quantum 4-dimensional Minkowski space-time with,
{\it via the gauge principle,} the force-fields
gauge-group structure $SU_c(3) \times SU_L(2)
\times U(1) \times SU_f(3)$ built-in from the
very beginning. This is the "background" of
everything, i.e., the lepton world and the
quark world. For notations, we use Wu and
Hwang \cite{Book}.

The gauge symmetry $SU_L(2) \times U(1) \times SU_f(3)$
applies to the lepton world, since the lepton world does
not recognize the strong interaction $SU_c(3)$. The
gauge symmetry $SU_c(3) \times SU_L(2) \times U(1)$ applies
to the quark world, and the "quark" group $SU_Q(3)$ (which
is not a gauge symmetry) leaves the mixing heavily among
the three "generations" of quarks. Note that one $SU(3)$
is enough but two of them would be too much, as far as
the gauge principle is concerned. {\it Thus, the deep
mystery can be resolved by introducing the "quark"
group $SU_Q(3)$, used to be called "the flavor $SU(3)$
symmetry". The quark entities are members of this group.}

In the quark world, we have the "basic unit of matter",
for the up-type right-handed quarks $(u_R,\, c_R,\,
t_R)$ (column), a triplet under the quark group $SU_Q(3)$,
\begin{equation}
D_\mu = \partial_\mu - i g_c {\lambda^a\over 2} G_\mu^a -
i {2\over 3} g'B_\mu,
\end{equation}
and, for the rotated down-type right-handed quarks $(d'_R,\, s'_R,
\, b'_R)$ (column), a triplet under the quark group $SU_Q(3)$,
\begin{equation}
D_\mu = \partial_\mu - i g_c {\lambda^a\over 2} G_\mu^a -
i (-{1\over 3}) g' B_\mu.
\end{equation}

On the other hand, we have the basic unit of matter, for the
$SU_L(2)$ quark doublets and a triplet under the quark
group $SU_Q(3)$,
\begin{equation}
D_\mu = \partial_\mu - i g_c {\lambda^a\over 2} G_\mu^a - i g
{\vec \tau\over 2}\cdot \vec A_\mu - i {1\over 6} g'B_\mu.
\end{equation}

{\it Thus, the term of "generation" is a misnomer when
the quark group $SU_Q(3)$ is introduced. Similarly,
the "generation" in the lepton world becomes a
misnomer after the gauge group $SU_f(3)$ is introduced.
On the quark group $SU_Q(3)$, we should recall the
role of the $CKM$ matrix \cite{CKM} to ascertain
that the elements of the quark group $SU_Q(3)$ should
be the input entities, rather that each individual
quark, in these discussions.}

At this juncture, we wish to discuss the basic role
of the gauge principle, a strange topic which was
discussed only along one line or two in the 20th
Century \cite{Book}. The reasons which we have to
decide the $D_\mu$ in the substitution $\partial_\mu
\to D_\mu$ for each basic unit of matter are
at least two-fold.

For each basic unit matter, such as the three
quark units as given in the above, there is only
one kinetic-energy term in the lagrangian, and
there should be only one.

The purpose of the kinetic-energy term is to tell
how the basic unit of matter changes when the
space-time point moves from one point to another.
When we introduce the quantum 4-dimensional
Minkowski space-time, the coordinates $(x_\mu)$
alone are not adequate to describe the motion of
the particle.

In the above example of the quark world, the
gauge principle requires the knowledge of the
force fields through the gauge fields. That is
why the gauge-group structure has to be implemented
{\it "from the very beginning"}.

Moreover, the implementation of $SU_c(3) \times
SU_L(2) \times U(1)$ precludes the assignment
of $SU_{family}(3)$ of the various basic units of
matter in the quark world. We presume the reason
might be that the $SU_f(3)$ coupling $\kappa$ is
much weaker than the strong $SU_c(3)$ coupling $g_c$.
Thus, the implementation of $SU_c(3)$ precludes
the contradictory assignments of $SU_f(3)$ from
occurrence. But the space-time itself is $SU_c(3)
\times SU_L(2) \times U(1) \times SU_f(3)$ in nature.

\medskip

As for the lepton world, we introduce the family triplets,
$(\nu_\tau^R,\,\nu_\mu^R,\,\nu_e^R)$ (column) ($\equiv \Psi_R(3,1)$)
and $(\tau_R,\,\mu_R,\,e_R)$ (column) ($\equiv \Psi_R^C(3,1)$),
under $SU_f(3)$. The gauge principle for the "basic unit of matter"
is, for $\Psi_R(3,1)$,
\begin{equation}
D_\mu = \partial_\mu - i \kappa {\bar\lambda^a\over 2} F_\mu^a.
\end{equation}
Similarly for $\Psi_R^C(3,1)$.

And, for the left-handed $SU_f(3)$-triplet and $SU_L(2)$-doublet
$((\nu_\tau^L,\,\tau^L),\, (\nu_\mu^L,\,\mu^L),\, (\nu_e^L,\,e^L))$
(all columns) ($\equiv \Psi_L(3,2)$), the basic unit is
\begin{equation}
D_\mu = \partial_\mu - i \kappa {\bar\lambda^a\over 2} F_\mu^a - i g
{\vec \tau\over 2} \cdot \vec A_\mu + i {1\over 2} g' B_\mu.
\end{equation}

The generation of the various quark masses is through the Standard-Model
way. For the charged leptons, we have to choose $\bar\Psi_L(\bar 3,2)
\Psi_R^C(3,1) \Phi(1,2)$. Note that, in
the U-gauge, the SM Higgs looks like $(0,\, {1\over \sqrt 2} (v+\eta(x) ))$,
only one degree of freedom \cite{Book}. Thus, the neutral neutrino triplet
$\Psi_R(3,1)$ cannot be used in the above coupling, because of the charge
conservation.

Moreover, there is an important point in the lepton world.
In the $SU_c(3) \times SU_L(2) \times U(1) \times SU_f(3)$ Standard Model
\cite{Hwang417}, the neutrino mass term assumes a new form:
\begin{equation}
i {h\over 2} {\bar\Psi}_L(3,2) \times \Psi_R(3,1) \cdot {\tilde \Phi}(3,2)
+ h.c.,
\end{equation}
where $\Psi(3,i)$ is the neutrino triplet (with
the first label for $SU_f(3)$ and the second for $SU_L(2)$). The
cross-dot (curl-dot) product is somewhat new, referring to the singlet
combination of three triplets in $SU(3)$. The Higgs field $\Phi(3,2)$ is
new in this effort \cite{HwangYan}, because it carries some nontrivial
$SU_L(2)$ charge. The consistent definition of ${\tilde Phi}(3,2)$ is
given below.

This off-diagonal neutrino mass term offers us a natural way to describe
neutrino oscillations, since the neutral part of $\Phi(3,2)$ could each
receive vacuum expectation values.

Note that, for charged leptons, the Standard-Model choice is ${\bar \Psi}_L(\bar 3,2)
\Psi_R^C(3,1) \Phi(1,2) +c.c.$, which gives three leptons an equal mass. But, in
view of that if $(\phi_1,\phi_2)$ is an $SU(2)$ doublet then $(\phi_2^\dagger,
-\phi_1^\dagger)$ is another doublet, we could form ${\tilde\Phi}^\dagger(3,2)$
from the doublet-triplet $\Phi(3,2)$. So, this $\Phi(3,2)$ is used below, in
a consistent notation as $\Phi(1,2)$, while its "related" ${\tilde \Phi}(3,2)$
is used above.

\begin{equation}
i {h^C\over 2} {\bar\Psi}_L(3,2) \times \Psi_R^C(3,1) \cdot
\Phi(3,2) + h.c..
\end{equation}
Here vacuum expectation values of $\Phi^0(3,2)$ give rise to
the imaginary off-diagonal (hermitian) elements in the
$3\times 3$ mass matrix, so removing the equal masses of the
charged leptons.

Here the couplings $h^C$ and $h$ are closely related to
the coupling strength $\kappa$ for the family gauge bosons
\cite{Family}. Note that all these couplings are dimensionless
in the 4-dimensional Minkowski space-time; thus, it is the
intrinsic property of the 4-dimensional Minkowski space-time
- in other words, it should be {\it not} adjustable by those
living in this space-time, at will.

The triple quark world in the quantum 4-dimensional Minkowski
space-time is governed by the six dimensionless
couplings, three for the strengths of the interactions
and another three mass-related parameters.
Similarly, the lepton world in the quantum 4-dimensional
Minkowski space-time is governed by six dimensionless
couplings, three for the interactions and three for the
mass-related parameters. The entries for the quark world
are members of the quark group $SU_Q(3)$ while the entries
for the lepton world are singlets under $SU_Q(3)$. Using
our arguments, the dimensionless couplings are determined
{\it globally} by the quantum 4-dimensional Minkowski
space-time, the covering space of the game. All these
might lead to some philosophical insights.

{\it Mathematics-wise, the Standard Model \cite{definition}
is basically a group theory, though fairly complicated.
The entry point is characterized by the basic units of
matter, which are expressed in terms of quarks,
electrons, neutrinos, etc. The gauge principle,
i.e., $\partial_\mu \to D_\mu$ which varies with the
basic unit of matter, also brings in the various gauge
fields (i.e., the force fields). The complex scalar
fields enter for the massive gauge fields. As the
beginning, there is a quark group $SU_Q(3)$ for which
the basic units of matter for quarks are triplets
while the basic units of matter for leptons are
singlets. The basic units of matter are described
by the Dirac equations of the anti-commuting and
closed 16-elements Dirac algebra. If we count
everything carefully, all physics, including the
hadron $CP$-violating phase, neutrino oscillations,
etc., are taken into account.}

\medskip

\section{The Origin of Mass}

As pointed out in an early version of the paper
\cite{Hwang417}, we may imagine that, in the U-gauge,
the Standard-Model Higgs $\Phi(1,2)$ looks like $(0,\,
(v+\eta(x))/\sqrt 2)\,\, (column)$ and
$\Phi^\dagger(\bar 3,2) \Phi(1,2)$ would pick out
the neutral sector naturally. In fact, the quartic
term $(\Phi^\dagger(\bar 3,2) \Phi(1,2))(\Phi^\dagger(1,2)
\Phi(3,2))$ with a suitable sign, would modify the mass
term for $\Phi(3,2)$ field such that the neutral sector
has SSB while the charged sector remains massive. This
"projected-out" Higgs mechanism is what we need.

We may \cite{Hwang417} write down the terms for potentials among the three
Higgs fields, subject to (1) that they are renormalizable, and (2) that
symmetries are only broken spontaneously (via the Higgs or induced Higgs
mechanism). Thus, we write, from the earliest (17 April 2013) version of
\cite{Hwang417},

\begin{eqnarray}
V = & V_{SM} +  V_1 + V_2 + V_3,\\
V_{SM} =& \mu^2 \Phi^\dagger(1,2) \Phi(1,2) +\lambda (\Phi^\dagger(1,2)
\Phi(1,2))^2,\\
V_1 =& M^2 \Phi^\dagger(\bar 3,2) \Phi(3,2) + \lambda_1
(\Phi^\dagger(\bar 3,2) \Phi(3,2))^2\nonumber\\
  & +\epsilon_1(\Phi^\dagger(\bar 3,2)\Phi(3,2))(\Phi^\dagger(1,2)\Phi(1,2))
    +\eta_1 (\Phi^\dagger(\bar 3,2)\Phi(1,2))(\Phi^\dagger(1,2)\Phi(3,2))
  \nonumber\\
  &+ \epsilon_2(\Phi^\dagger(\bar 3,2)\Phi(3,2))(\Phi^\dagger(\bar 3,1)\Phi(3,1))
   + \eta_2 (\Phi^\dagger(\bar 3,2)\Phi(3,1))(\Phi^\dagger(\bar 3,1)\Phi(3,2))
   \nonumber\\
  &+ (\delta_1 i \Phi^\dagger(3,2) \times \Phi(3,2) \cdot \Phi^\dagger(3,1)
  + h.c.),\\
V_2 =& \mu_2^2 \Phi^\dagger(\bar 3,1)\Phi(3,1) + \lambda_2
(\Phi^\dagger(\bar 3,1) \Phi(3,1))^2 + (\delta_2 i \Phi^\dagger(3,1)\cdot
\Phi(3,1) \times \Phi(3,1) +h.c.)\nonumber\\
 &+ \lambda_2^\prime\Phi^\dagger(\bar 3,1) \Phi(3,1) \Phi^\dagger(1,2) \Phi(1,2),\\
V_3 =& (\delta_3 i \Phi^\dagger(3,2)\cdot \Phi(3,2)\times
    (\Phi^\dagger(1,2)\Phi(3,2))+h.c.)\nonumber\\
    & +(\delta_4 i (\Phi^\dagger(3,2)\Phi(1,2))\cdot \Phi^\dagger(3,1)
    \times \Phi(3,1)+h.c.)\nonumber\\
    & + \eta_3 (\Phi^\dagger(\bar 3,2)\Phi(1,2)\Phi(3,1)+c.c.).
\end{eqnarray}
In obtaining the last expression, we maintain that every terms are naively
renormalizable since the terms are of power four or less (in scalar
fields). Note that the terms in $\delta_j i$ involve the so-called
"$SU(3)$ operations", as before.

This is {\it the first step} to decide what the Standard
Model is going to be.

As shown earlier \cite{Family, HwangWYP}, two triplets of complex scalar
fields would make the eight family gauge bosons and four residual family
Higgs particles all massive. In the $SU_f(3)$ family gauge theory
treated alone, there are many ways of accomplishing such goal. In
the present complicated case, the equivalence between two triplets is
lost, but we discover that this modification would be a nice way to
patch up that three neutrinos couple to independent vacuum expectation
values (and the family Higgs), explaining neutrino oscillations among
three generations.

The "principle" of renormalizability turns out to be very powerful. To
the least, we may use it to define the calculability of the theory. It
implies the existence of some branch of calculable mathematics (in
quantum field theory).

On the other hand, there are too many terms - for example, there
are three "ignition" terms; if there is only one, we must make
the smart choice; etc. If we look into the lepton world or into
the quark world, it is striking to realize that all the couplings
are dimensionless in the 4-dimensional Minkowski space-time. In the
Higgs and gauge sector, we could take only one "ignition" term
and take all the rest couplings to be dimensionless - then, the
Standard Model is virtually a complete dimensionless theory (here
"virtually" means "apart from the 'ignition' term").

This is {\it the second (last) step} to decide the
Standard Model is going to be - to figure out what
the dimensionless theory looks like.

Thus, "renormalizability" plus "dimensionless-ness in
the 4-dimensional Minkowski space-time" leads to the
final theory - is that elegant?

Before 2014, our understanding of mass generation for the building blocks of
matter was still lacking, not mentioning the origin of mass; so, the same
story for the mysterious Higgs mechanisms. The story has changed completely
if we accept the recent proposal \cite{Origin} on the origin of mass.

The Higgs lagrangian (i.e., Eqs. (8)-(12) above) is based on
renormalizability. Physics-wise, we should think of its
redundancy - for example, if there is a spontaneous symmetry
breaking (SSB), there are three different channels for
switch-on, depending on which complex scalar field gets
turn on. On the other hand, when the temperature is high
enough, those mass terms, except the switch-on of ignition,
would become negligible; or, before the generation of
masses.

Thus, the real world should be like this: There is at most
one ignition channel for spontaneous symmetry breaking (SSB).
When the temperature is high enough, all the mass terms should
be zero (if necessary, by comparison).

In the 4-dimensional Minkowski space-time, the complex scalar
field $\phi(x)$ have one unique feature - that the renormalizable
term $\lambda (\phi^\dagger \phi)^2$ is allowed (only in the
4-dimensional space-time) \cite{definition}. This self-interaction
is repulsive, explaining why we cannot see it unless under
exceptions. Further, this term is dimensionless, so the
coupling $\lambda$ is a pure number that is determined by
the 4-dimensional Minkowski space-time, {\it not} by the
complex scalar field itself \cite{Origin, definition}.

Thus, one important beginning point is the 4-dimensional
Minkowski space-time. The other point is whether the
force-fields gauge symmetry $SU_c(3) \times SU_L(2) \times
U(1) \times SU_f(3)$ should also be a part of this
beginning point. As all the future entities should
have these gauge group assignments (similar to quantum
numbers), our suggestion is "yes" - that gives the assertion
that "we live in the 4-dimensional Minkowski space-time
with the force-fields gauge group $SU_c(3) \times SU_L(2)
\times U(1) \times SU_f(3)$ built-in from the outset"
\cite{definition}.

Let us re-iterate the origin of mass \cite{Origin}:

Suppose that, before the spontaneous symmetry breaking (SSB), the Standard Model
does not contain any parameter that is pertaining to "mass", but, after the SSB,
all particles in the Standard Model acquire the mass terms as it should - we
call it "the origin of mass". In \cite{Origin}, we show that this is indeed
the case - explaining the origin of mass. In this way, we sort of tie "the
origin of mass" to the effects of the SSB, or the generalized Higgs mechanism.

In our opinion, this way of explaining the origin of mass is of fundamental
importance. After all, we could not give some real meaning to the term "mass";
it makes sense to interpret masses as our proposal of the origin of
mass \cite{Origin}. According to \cite{Origin}, the masses would disappear
altogether at the temperature high enough, if for example we are thinking
of the early Universe. On other hand, when the matter system becomes so
dense and so hot in the process of forming a "black hole", the mass
parameters would lose its meaning altogether - making the interior inside the
horizon of the Schwarzschild metric something which we can't understand so
far.

Starting from the Higgs sector of the Standard Model \cite{Hwang417}, we
have the Standard-Model Higgs $\Phi(1,2)$, the purely family Higgs
$\Phi(3,1)$, and the mixed family Higgs $\Phi(3,2)$, with the first label
for $SU_f(3)$ and the second for $SU_L(2)$. We need another triplet $\Phi(3,1)$
since all eight family gauge bosons have to be massive \cite{Family}.

We could say that it is the most interesting to put the different but related
complex scalar fields (Higgs) in the same pot, letting them interact together.
Two fields $\phi_a$ and $\phi_b$ can interact "directly" via
$(\phi_a^\dagger\phi_a)(\phi_b^\dagger\phi_b)$ but, if related,
via $((\phi_a^\dagger\phi_b)(\phi_b^\dagger\phi_a))$. Here we
always talk about those things which are renormalizable. Under the
renormalizable game, these are closely linked with the quartic
self coupling such as $\lambda (\phi_a^\dagger\phi_a)^2$.

Besides, in the 4-dimensional Minkowski space-time, these terms
have dimensionless couplings, or pure numbers. So, $\lambda$
depends on the 4-dimensional Minkowski space-time, {\it not} on
the field $\phi(x)$ itself. This is a very interesting aspect.
{\it Thus, when we put three complex scalar fields $\Phi(1,2)$,
$\Phi(3,2)$, and $\Phi(3,1)$ in this pot of the 4-dimensional
Minkowski space-time all the quartic terms, including those
crossed, are playing pure-number games among themselves.
Consequently, the masses for the various Higgs emerge as the
result. This discovery is closely linked to the origin of fields
(point-like particles) \cite{definition}.}

In fact we are "playing" with "energies", in terms of the
complex scalar fields. For two related complex scalar
fields, the cross term helps to lower the total energy.
For two unrelated complex scalar fields, the cross
term does not exists.

In the U-gauge, we choose to have
\begin{equation}
\Phi(1,2)= (0,{1\over \sqrt 2} (v+\eta)),\,\, \Phi^0(3,2) = {1\over \sqrt 2}
(u_1+\eta'_1, u_2+ \eta'_2, u_3+\eta'_3 ),\,\, \Phi(3,1) = {1\over \sqrt 2}(w+\eta',0,0),
\end{equation}
all in columns. The five components of the complex triplet $\Phi(3,1)$ get
absorbed by the $SU_f(3)$ family gauge bosons and the neutral part of
$\Phi(3,2)$ has three real parts left - together making all eight family
gauge bosons massive.

Before the mixing, the masses of the various Higgs are given by, using the general
Higgs lagrangian (cf. Eqs. (8)-(12)),
\begin{eqnarray}
\eta:& (\mu^2/\lambda) + {1\over 4}(\epsilon_1+ \eta_1) u_i u_i + {\lambda'_2\over 4}
w^2, \nonumber\\
\eta':& (\mu_2^2/\lambda_2) + {\epsilon_2\over 4} u_i u_i +{\eta_2\over 4} u_1^2 +
{\lambda'_2\over 4} v^2,
\nonumber\\
\eta'_1:& M^2 + {1\over 4}(\epsilon_1+ \eta_1) v^2 +{\epsilon_2\over 4} w^2 +
(\lambda_1-term),\nonumber\\
\eta'_{2,3}:& M^2 + {1\over 4}(\epsilon_1+ \eta_1) v^2 + {\epsilon_2\over 4} w^2
+ (\lambda_1-term),\nonumber\\
\phi_1:& M^2 + {1\over 2}\epsilon_1 v^2 + {1\over 2}\epsilon_2 w^2 + {1\over 2}\eta_2 w^2 +
{\lambda_1\over 2} u_i u_i, \nonumber\\
\phi_{2,3}:& M^2 + {1\over 2}\epsilon_1 v^2 + {1\over 2} \epsilon_2 w^2
+ {\lambda_1 \over 2} u_i u_i.
\end{eqnarray}
The mixing term looks like, apart from some common factor:
\begin{equation}
2 (\epsilon_1+\eta_1)u_i\eta'_i v \eta + 2\epsilon_2 u_i\eta'_i w \eta' +
2 \eta_2 u_1\eta'_1 w\eta' + 2\lambda'_2 w\eta' v\eta.
\end{equation}

There are the other mixings (such the mixing inside
$\eta'_{1,2,3}$), to be discussed later in this
paper. To work out on "the origin of mass", we would
drop out all "mass" terms to begin with.

Let us try to fix the notations further. See Ch. 13,
Ref. \cite{Books}. For $\eta'$ going through SSB, we have the following
terms, neglecting the small ${\lambda'}_2$ term,
\begin{equation}
{\mu_2^2\over 2}(\eta'+w)^2+({\epsilon_2\over 4} u_i u_i +{\eta\over 4}
u_1^2) (\eta'+w)^2 + {\lambda_2\over 4}(\eta'+w)^4. \nonumber\\
\end{equation}
SSB means that all the linear terms add up to zero, resulting the
change in sign of the mass term, ${1\over 2} (2\lambda_2 w^2 (\eta')^2)$.
Note that, for real fields, a factor of ${1\over 2}$ should be factored
out.

The same applies to other SSB fields, even though the original
minus signs are generated by other fields. This applies for the
SM Higgs $\eta$.

In other word, the three "related" scalar (Higgs) fields
$\Phi(1,2)$, $\Phi(3,2)$, and $\Phi(3,1)$ should be "equivalent"
among themselves. The spontaneous symmetry breaking (SSB) is
happening for all of them, actively or passively. Thus,
$\lambda$, $\lambda_1$, and $\lambda_2$ should be the same.
$\lambda'_2$ should be zero or vanishingly small.

All the mass terms, including those SSB-driving terms, should
be absent, apart from one SSB-driving term. The scalar field ($\phi_a$)
could affect a different scalar field ($\phi_b$), if they
are triplets under $SU_f(3)$, etc. It is easy to see that only
one SSB-driving term is enough for all the three Higgs fields --
there may be several SSB's for the neutral fields - in our case,
it works for all of them.

We are working with the lagrangian, i.e., the energy that should
be bounded from below ("positive definiteness"). The combination
$(\phi^\dagger_a\phi_a+\phi^\dagger_b\phi_2)^2 - 4 (\phi^\dagger_a
\phi_b)(\phi^\dagger_b \phi_a)$ would work very well. This in fact
helps to fix $\epsilon_{1,2}$ and $\eta_{1,2}$. As we argued before,
the three $\lambda$'s are the same.

As for the SSB-driving term, we decide to keep the purely
family term, $\mu^2_2 \Phi^\dagger(3,1)\Phi(3,1)$. As we shall
see that the symmetry breaking would occur much earlier (in the
history of the early Universe), or at a much higher temperature.

Thus, we have

\begin{eqnarray}
V_{Higgs} =& \mu^2_2 \Phi^\dagger(3,1) \Phi(3,1) + \lambda
(\Phi^\dagger(1,2) \Phi(1,2)+ cos\theta_P\Phi^\dagger(3,2)\Phi(3,2))^2\nonumber\\
    &  + \lambda(-4 cos\theta_P)
(\Phi^\dagger(\bar 3,2)\Phi(1,2))(\Phi^\dagger(1,2)\Phi(3,2))
  \nonumber\\
  &+\lambda
(\Phi^\dagger(3,1) \Phi(3,1)+ sin\theta_P \Phi^\dagger(3,2)\Phi(3,2))^2
\nonumber\\ & + \lambda(-4 sin\theta_P)
(\Phi^\dagger(\bar 3,2)\Phi(3,1))(\Phi^\dagger(3,1)\Phi(3,2)).
\end{eqnarray}
These are two perfect squares minus the other extremes, to guarantee
the positive definiteness, when the minus $\mu^2_2$ was left out.
($\theta_P$ may be referred to as "Pauchy's angle".) $\epsilon_1$,
$\eta_1$, and $\epsilon_2$, $\eta_2$ are expressed in $\lambda$, a
great simplification. Note that we only include the interference
terms between those involving the same group, $SU_f(3)$ or $SU_L(2)$;
thus $\lambda'_2 =0$.

From the expressions of $u_iu_i$ and $v^2$, we obtain
\begin{equation}
v^2 (3 cos^2\theta_P-1) = sin\theta_P cos\theta_P w^2.
\end{equation}
And the SSB-driven $\eta'$ yields
\begin{equation}
w^2 (1-2 sin^2\theta_P) = - {\mu_2^2\over \lambda} +
(sin2\theta_P - tan\theta_P) v^2.
\end{equation}
These two equations show that it is necessary to have the driving
term, since $\mu^2_2=0$ implies that everything is zero. Also,
$\theta=45^\circ$ is the (lower) limit.

The mass squared of the SM Higgs $\eta$ is $2\lambda cos\theta_P u_i u_i$,
as known to be $(125\,\, GeV)^2$. The famous $v^2$ is the number
divided by $2\lambda$, or $(125\,\,GeV)^2/(2\lambda)$. Using PDG's for
$e$, $sin^2\theta_W$, and the $W$-mass \cite{PDG}, we find
$v^2=255\,\, GeV$. So, $\lambda={1\over 8}$, a simple model indeed.

The ratio of the VEV to its Higgs mass is determined
by $2\lambda$, whether the channel is not ignited or not. We might
choose the channel of $\eta'$ (the purely family Higgs) or that of
$\eta$ (the SM Higgs) as the ignition channel, but three Higgs channels
have different labels. The three Lorentz-invariant scalar fields have
different internal structures - an amusing question for further
investigation.

The mass squared of $\eta'$ is $-2(\mu_2^2-sin\theta_P u_1^2 +
sin\theta_P (u_2^2+u_3^2))$. The  other condensates are $u_1^2= cos\theta_P v^2
+ sin\theta_P w^2$ and $u_{2,3}^2 = cos\theta_P v^2 - sin\theta_P w^2$ while
the mass squared of $\eta'_1$ is $2\lambda u_1^2$, those of $\eta'_{2,3}$ be
$2\lambda u_{2,3}^2$. The mixings among $\eta'_i$ themselves are neglected
in this paper.

There is no SSB for the charged Higgs $\Phi^+(3,2)$. The mass
squared of $\phi_1$ is $\lambda(cos\theta_P v^2 - sin\theta_P w^2) + {\lambda\over 2}
u_i u_i$ while $\phi_{2,3}$ be $\lambda(cos\theta_P v^2 + sin\theta_P w^2)
+ {\lambda\over 2} u_i u_i$. (Note that a factor of ${1\over 2}$ appears
in the kinetic and mass terms when we simplify from the complex case to
that of the real field; see Ch. 13 of \cite{Book}.)

A further look of these equations tells that $3cos^2\theta_P - 1 > 0$ and
$2sin^2\theta_P -1 > 0$. A narrow range of $\theta_P$ is allowed (greater
than $45^\circ$ while less than $57.4^\circ$, which is determined by
the group structure). For illustration, let us choose
$cos \theta_0 = 0.6$ and work out the numbers as follows:
(Note that $\lambda={1\over 8}$ is used.)
\begin{eqnarray}
& 6 w^2 = v^2, \quad -\mu^2_2/\lambda = 0.32 v^2;\nonumber\\
\eta: & 2\lambda cos\theta_0 u_i u_i =(125\, GeV)^2, \quad v^2 = (250\,GeV)^2;
\nonumber\\
\eta': & mass^2 = (51.03\,GeV)^2, \quad w^2=v^2/6; \nonumber\\
\eta'_1: & mass^2= (107\,GeV)^2, \quad u_1^2=0.7333 v^2; \nonumber\\
\eta'_{2,3}: & mass^2 = (85.4\,GeV)^2, \quad u_{2,3} = 0.4667 v^2; \nonumber\\
\phi_1:& mass = 100.8\, GeV; \qquad \phi_{2,3}: mass = 110.6\,GeV.
\end{eqnarray}
All numbers appear to be reasonable. In the above, $cos\theta_0$ is the
only free parameter until one of the family Higgs particles $\eta'_{1,2,3}$
and $\eta'$ is found experimentally. Since the new objects need to be
accessed in the lepton world, it would be a challenge for our experimental
colleagues.

{\it As a footnote, our Standard Model predicts that the mass of the SM
Higgs $\eta$ is a half of the vacuum expectation value $v$ - a prediction
in the origin of mass \cite{Origin}.}

As for the range of validity, ${1\over 3} \le cos^2\theta_P \le {1\over 2}$.
The first limit refers to $w^2=0$ while the second for $\mu_2^2 = 0$.

We may fix up the various couplings, using our common senses. The
cross-dot products would be similar to $\kappa$, the basic coupling of
the family gauge bosons. The electroweak coupling $g$ is
$0.6300$ while the strong QCD coupling $g_s=3.545$; my first guess
for $\kappa$ would be about $0.1$. The masses of the family gauge
bosons would be estimated by using ${1\over 2}\kappa \cdot w$, so
slightly less than $10\,GeV$. (In the numerical example with $cos
\theta_P=0.6$, we have $6 w^2= v^2$ or $w=102\,\,GeV$. This gives
$m=5\,\,GeV$ as the estimate.) So, the range of the family forces,
existing in the lepton world, would be $0.02\,\, fermi$.

Furthermore, the $\lambda'_2$ term (i.e., the interference between
the spaces (3, 1) and (1, 2)) could be neglected safely in these
discussions. But similar discussions to justify our Standard Model
could be presented. There is one interesting aspect that the scalar
fields, when having same $SU_f(3)$ or $SU_L(2)$ group structure,
should be dealt with together, like in this paper.

The term that ignites the SSB is chosen to be with
$\eta'$, the purely family Higgs. This in turn ignites EW SSB
and others. It explains the origin of all the masses, in terms
of the spontaneous symmetry breaking (SSB). SSB in $\Phi(3,2)$
is driven by $\Phi(3,1)$, while SSB in $\Phi(1,2)$ from the
driven SSB by $\Phi(3,2)$, as well. The different, but related,
scalar fields can accomplish so much, to our surprise.

There is more reason that we could be optimistic. Besides the $SU_c(3)$ protection
over the quark world, we now have another $SU_f(3)$ to protect the lepton world -
the remaining part of the world in the Standard Model. The (conceptual) troubling
QED Landau ghost is no longer there since QED alone is only part of the story;
it is all asymptotically free, via $SU(3)$, and free of the ghosts. Maybe
everything can be put together elegantly as a final complete and consistent
theory.

On the experimental side, the family Higgs $\eta'_1$,
$\eta'_{2,3}$, and $\phi^+_{1,2,3}$ couple to only
cross-generation leptons - making them difficult to
observe. On the other hand, the purely family Higgs
$\eta'$ is very elusive - maybe hopeless to observe.
It is an ideal candidate particle for the so-called
dark matter. These all happen in the lepton world,
if the Standard Model (of this paper) is valid.

In the (theoretical) processes of figuring out how to write
the Standard Model, or what is the best way to write it
down, which would be the first to spell out (i.e. the
"background"), etc., it came out with "a very precise
definition of the Standard Model" \cite{definition}
and then the word "built-in from the outset"
\cite{definition}. It is important to figure out these
steps in detail - it should eventually let us
understand the Nature.

Half a year before T.-M. Yan and I wrote the paper
\cite{HwangYan} to put six objects in the multiplet
$\Psi_L(3,2)$, I was rather reluctant to do so even
though I did have some important breakthroughs
\cite{Hwang10} that eventually drove me till the
end. One should remember that the
processes of searching for the Standard Model is
non-orthodox, varying from a physicist to another.
To be honest, we are not sure that a specific
Standard Model would survive, or would
survive for how long. It is just amazing that the
entire body of particle physics, or the Nature,
could be summarized by the Standard Model -
the description of the {\it smallest} units of
matter, such as electrons, neutrinos, and quarks,
on the basis of Einstein's relativity principle
and the quantum principle.

\medskip

\section{The Statement of the Standard Model}

To describe the motion of matter, we have to add the
knowledge of the velocities on top of the space-time
coordinates, according to Newton's doctrine.
The gauge principle says $\partial_\mu \to D_\mu$
with $D_\mu$ containing the gauge fields. Thus, the
information on the gauge fields should be given,
{\it a priori}, the knowledge of the space-time
coordinates.

To be precise, thus, we could state the Standard
Model as follows:

"We should declare that we are living in
the quantum 4-dimensional Minkowski space-time
with, via the gauge principle, the force-fields
gauge-group structure $SU_c(3) \times SU_L(2)
\times U(1)$ (for the quark world) or $SU_L(2)
\times U(1) \times SU_f(3)$ (for the lepton
world) built-in from the very beginning.
The input entries are members of the quark
group $SU_Q(3)$, the so-called "the triple
quark world" and each of them enters in a
way both dimensionless and massless, of
nuclear sizes and of the $SU_c(3)\times
SU_L(2) \times U(1)$ gauge symmetry.
Meanwhile, the lepton world is dimensionless
and massless, of atomic sizes and of the
$SU_L(2) \times U(1) \times SU_f(3)$ gauge
symmetry."

The (123) gauge symmetry, the symmetry under
$SU_c(3) \times SU_L(2) \times U(1)$, means
that there exist $SU_Q(3)$ quark triplets,
$\Psi_L(3,2,1)$, $\Psi_R^u(3,1,1)$, and
$\Psi_R^{d'}(3,1,1)$, as the Dirac entry point.
The other (123) gauge symmetry, the symmetry
under $SU_L(2) \times U(1) \times SU_f(3)$,
means that there exist $SU_Q(3)$ lepton
singlets, $\Psi_L(1,2,3)$, $\Psi_R(1,1,3)$,
and $\Psi_R^C(1,1,3)$, as another Dirac entry
point. Here the three numbers refer to the
assignments under the gauge groups: the
color group, the $SU_L(2)$, and the $SU_f(3)$.
Usually, we do not mention the assignment
under the color group, for the sake of
simplicity.

Mathematically, the Standard Model
\cite{definition} is a group theory - a
group of elements, triplets and singlets,
of the quark group $SU_Q(3)$; they interact
among themselves through the symmetry of
the gauge group, either $SU_c(3) \times
SU_L(2) \times U(1)$ or $SU_L(2) \times
U(1) \times SU_f(3)$; they play out on
the background of the quantum
4-dimensional Minkowski space-time.

We remark that the meaning of the so-called
"generations" disappears altogether in this
(new) Standard Model. The input of the quark
group $SU_Q(3)$ and the introduction of the
gauge symmetry $SU_f(3)$ kill the old notion
of the "generations".

Even though both the triple quark world and
the lepton world are dimensionless, the
length scales, $10^{-13} cm$ versus
$10^{-8} cm$, still have the vastly
difference - one mystery of the quantum
4-dimensional Minkowski space-time.

First, we should remark on the gauge principle,
of which the meaning could be rather obscure
previously but, now, may be clarified. The
substitution, $\partial_\mu \to D_\mu$, for
each smallest unit of matter is required when
the space-time coordinates $x_\mu$ move to
a neighboring point. Thus, it has to be
built-in from the very beginning. {\it The
gauge principle is a fundamental mathematical
principle.}

Second, renormalizability should be promoted
to "the principle of renormalizabily". Not just
we should include $\lambda(\phi^\dagger \phi)^2$
for the complex scalar field $\phi(x)$, but also
{\it all} other renormalizable terms, including
the interactions among different fields, should
be included. Again, the principle of
renormalizability should be meaningful in the
sense of mathematics.

Thirdly, the structure of the Standard Model
\cite{definition} should be noted: Apart from
the "ignition" term, all the couplings are
dimensionless in the 4-dimensional Minkowski
space-time. {\it It is a dimensionless
theory in the quantum 4-dimensional Minkowski
space-time, if we could ignore the "ignition" term.}
It is a beautiful theory.

Lastly, the previous treatments of ultraviolet
divergences all were based on the massive
theories, resulting in divergences in
$ln ({M^2\over m^2})$ or in $({M^2\over m^2})^n$,
etc. [See, for example, the references
\cite{definition, Book} for further
discussions.] But the Standard Model is a massless
theory apart from the $SSB$ "ignition" term. Maybe
this opens the door to solve this century-old puzzle.

We should challenge our young colleagues in carrying
out systematic calculations of these ultraviolet
divergences, as we don't feel easy toward why a beautiful
theory (the Standard Model) would contain these
diseases - maybe they are not there, or they are
there for good reasons. The physicists of the
20th Century got beaten by these ultraviolet
divergences, but those of the 21st Century may
write a different chapter of the history.

\medskip

\section{Family Interactions for Leptons}

The weak interactions involving leptons are determined
by $-{\bar \Psi}(3,2)\gamma_\mu D_\mu\Psi(3,2)$ [with $D_\mu$
given by Eq. (5)] plus two curl terms [Eqs. (6) and (7)].

For the basic processes such as the muon decay,
$\mu^- \to e^- + {\bar \nu}_e + \nu_\mu$, we may write,
symbolically, the transition amplitude \cite{Book}:
\begin{equation}
i T = {G_F \over \sqrt 2} {\bar u}_e(p') \gamma_\lambda(1+\gamma_5) v_e(k')
\cdot {\bar u}_{\nu_\mu} (k) \gamma_\lambda (1+\gamma_5) u_\mu(p) + others,
\end{equation}
but this might be incorrect, since $u(p)$, $v(k)$, etc. are, by definition,
on the mass shells. Neutrino oscillations tell us $u_{\nu_\mu}(k) \equiv
U_{\mu i} u_i (k)$, with the left-hand side defined by "$\equiv$";
similarly for the antineutrino $v_e(k)$.

In fact, our language here is only for the mass-shell Dirac
spinors, not for something which oscillates. So, we should
write $\sum_i U_{e i} u^i(k)$ for the electron-like neutrino,
etc., since $u^i(k)$'s are the mass-eigenstates - that is how
we set up the Dirac equations for.

Similarly, for the muon or the electron, they should be on mass
shells in our language - that is the way which we represent
the muon or the electron. So, this seems against the
off-diagonal mass terms - in fact, this requires the
overall consistency check at least.

Thus, we should write, for the muon decay,
\begin{eqnarray}
i T = & {G_F \over \sqrt 2} {\bar u}_e(p') \gamma_\lambda(1+\gamma_5) U_{e i} v_i(k')
\cdot {\bar u}_j (k) U^\dagger_{\mu j} \gamma_\lambda (1+\gamma_5) u_\mu(p) + \nonumber\\
  & {G' \over \sqrt 2} {\bar u}_e(p') (1-\gamma_5) \times U_{\mu j}v_j(k')
\cdot {\bar u}_i (k) U^\dagger_{e i} (1-\gamma_5) \times u_\mu(p).
\end{eqnarray}
Here the second term  is due to the charged family-Higgs exchange
$\phi_1$. (Our $\phi_1$ or $\eta'_1$ refers to the $\tau$ channel, by our
convention.) So, $G'$ [$\propto (h^C)^2/m(\phi^\dagger)^2$] is much smaller
than the Fermi coupling $G_F$. The $G'$ term arises from the product of
the two cross-dot products; the Fierz reordering could be used in this
context.

The important point is that all the Dirac spinors are
on the mass shells - in order that the expression can
be further calculated.

{\it The differential cross sections, or the decay rates, can easily
be calculated: The inference term between the dominant muon-decay
amplitude and the family-Higgs exchange term is relatively
tiny since the neutrino mass enters the numerator. The family-Higgs
exchange term, when squared, may be detectible if $h$ (or $\kappa$,
essentially) is not too small - here the mass of $\eta'_1$ is 107
GeV in the above numerical example. In other words, the
interference term is suppressed by the neutrino-mass effect
while the square of the family-Higgs exchange term could have
some observable effects. On a second thought, we should
classify the experimental efforts to search for effects
caused by the family Higgs in ordinary muon decays as
"the precision experiments".}

The other important example is the $\mu \to e$
conversions, say, $\mu^-(p) + p(q) \to e^-(p') + p(q')$.
The amplitude is given by
\begin{eqnarray}
i T = & {1\over (2\pi^4)} \int d^4 k {\bar u}_e(p') \cdot i
{1\over 2 \sqrt 2} {e\over sin\theta_W} \cdot i \gamma_\lambda
(1 +\gamma_5) \nonumber\\
&\cdot {1\over i} \sum_j U_{e j}^\dagger {m_j - i \gamma\cdot k
\over m_j^2 + k^2 - i\epsilon}\cdot i\cdot {i \over 2} h \cdot (-)
\cdot {1\over 2} (1- \gamma_5)\cdot u_i {1\over \sqrt 2}\cdot
\nonumber\\
&\cdot {1\over i} \sum_l {m_l -i \gamma\cdot k\over m_l^2 +k^2 -i\epsilon}
U_{\mu l} \cdot i {1\over 2\sqrt 2} {e\over sin\theta_W}\cdot
1 \gamma_\eta (1+ \gamma_5) \cdot u_\mu (p) \nonumber\\
&\cdot {\bar u} (q') \cdot i {1\over 2\sqrt 2} {e\over sin\theta_W}\cdot
i \gamma_\eta (1+\gamma_5)\cdot {1\over i} {m-i\gamma\cdot (q+k-p')
\over m^2 + (q+k-p')^2-i\epsilon}\nonumber\\
& \cdot i {1\over 2\sqrt 2} {e\over sin\theta_W}\cdot i\gamma_\lambda
(1+\gamma_5)\cdot u(q)\cdot {1\over i}{1\over m_W^2 +(k-p')^2 -i \epsilon}
\cdot {1\over i}{1\over m_W^2+(k+p)^2-i\epsilon},
\end{eqnarray}
in the 't Hooft-Feynman gauge. This means that the amplitude is
finite but not complete, but, as seen below, all are rather small.

Simplifying it further, we obtain
\begin{eqnarray}
i T = & ({1\over 2\sqrt 2}{e\over sin\theta_W})^4\cdot
i (-) {hu_1\over 2\sqrt 2}\cdot (-i\cdot 2)^2\nonumber\\
&\cdot {1\over (2\pi)^4} \int d^4k {\bar u}_e(p')\gamma_\lambda
\sum_{j,l} U^\dagger_{e j}\gamma\cdot k m_l U_{\mu l} \gamma_\eta
(1+\gamma_5)u_\mu(p) \nonumber\\
&\cdot {\bar u}(q') \gamma_\eta \gamma\cdot (q+k-p')\gamma_\lambda
(1+\gamma_5)u(q) \nonumber\\
& \cdot {1\over m_j^2+k^2-i\epsilon}\cdot {1\over m_l^2+k^2 -i\epsilon}
\cdot{1\over m^2+(q+k -p')^2-i\epsilon}\cdot{1\over m_W^2+(k-p')^2-i\epsilon}
\cdot{1\over m_W^2 + (k+p)^2-i\epsilon}.
\end{eqnarray}
Comparing this amplitude to the dominant muon-decay amplitude,
it is down by $u_1 m_l$ with the tiny neutrino mass $m_l$ (and
$u_1= 107 \, GeV$ in the above table) - about $10^{-9} GeV^2$.
The remaining reduction is from the 2nd-order weak interaction
- a factor of $10^{-5}$. So, normally, we expect $10^{-28}$ smaller
than the dominant modes.

We have discussed the $\mu \to e + \gamma$ before
\cite{HwangYan}. The decay rate is suppressed by the
neutrino mass effect $u_1 m_l$ and further by
gauge invariance (for a real photon). The mode turns
out to be smaller than the current search limits by
more than 20th order away (completely negligible).

To sum up, the modifications in the ordinary muon decays
may be observed - it defines a new category of the high-precision
experiments. The $\mu \to e$ conversions such as $\mu^- + p \to
e^- + p$ and $\mu \to e + \gamma$ turn to be hopelessly small.

Nevertheless, in certain decays such as $^3H \to ^3He + e^- +
{\bar \nu}_e$, we have to rewrite the neutrino state as a sum
of neutrino-mass eigen-states - as required by neutrino
oscillations. These are another category of high-precision
experiments.

{\it In a short summary, we do expect to see some corrections in
the ordinary muon decay $\mu^- \to e^- + {\bar \nu}_e + \nu_\mu$
but the branching ratios induced by the $\nu_\mu \to \nu_e$
conversion or vice versa, such as $p + \mu^- \to p+ e^-$ or
$\mu \to e + \gamma$, turn out to be rather small. In the muon
decay or nuclear beta decays, such as $^3H \to ^3He + e^- +
{\bar \nu}_e$, the replacement, as required by neutrino
oscillations, of flavor states by the mass eigen-states does
produce some tiny observable effects which might be detected
in the precision experiments of the next generation.}

\smallskip

Besides the three Higgs, the primary prediction of our Standard Model
is the existence of the force of a new kind - i.e., the family force
mediated by the family gauge bosons. As said above, we could use
${1\over 2}\kappa w$ as an estimate of the mass(es) of the family
gauge bosons. Our first guess is for some feeble force - $\kappa=0.1$.
The above numerical example corresponds to $w=102\,\, GeV$, so as to
the family gauge boson mass of $5\,\,GeV$.

The family gauge bosons would then be in the vicinity
of $5\,\, GeV$ or nearby, or the range of $0.04\,\,fermi$.
Or, $0.04 \times 10^{-13}$ cm in the
effective range, between leptons (such as two electrons or an
electron-positron pair) is too short to be detected
for the entire atomic physics or the entire chemistry.

The precision experiments such as $g-2$ would eventually detect the
residual family effects, since the existing $g-2$ calculations
\cite{Kinoshita} is so far the QED calculation and should be
completed by inclusion of other effects with the emphasis on
family gauge bosons. We are looking forward to prospects in this
direction.

Of course, we need to examine the precision part of atomic physics
when the story becomes clear; even though the effects are tiny,
the evolutions usually come from the tiny effects to begin with.

\medskip

\section{Experimental Questions yet to be Answered}

We are talking about the Standard Model \cite{definition}
in that there is a (new) force, conducted by the family
gauge bosons, that there are (new) family Higgs particles,
and that the existing smallest units of matter, including
electrons, neutrinos, and quarks, all of them are forming
the triplets and the singlets of the quark group $SU_Q(3)$
to begin with. Mathematically, it is a group theory - a
group theory of the Nature.

Surprisingly, the Standard Model \cite{Hwang417, definition}
is just a group theory, mathematically. The Standard Model
is used to describe our Universe. So, our Universe can be
described by, mathematically, a group theory. This
recognition of our Universe being described by a group
theory, mathematically, gives a lot of comforts to minds
of philosophers, or of mathematicians, or of theoretical
physicists.

This hypothetical fact should set the stage for testing
the complicated theory from the experimental ends.

There are no longer the three "generations" of leptons,
since they need something to make up entities of $SU_f(3)$.
There are no more the three "generations" of quarks,
since the basic entry entities are the triplets of the
quark group $SU_Q(3)$. The $CKM$ matrix \cite{CKM}
is naturally borne there with the triplets of $SU_Q(3)$.
The misnomer of "three generations" comes from lack of
the understandings in our previous knowledge.

The beauty of the Standard Model \cite{definition}
should prompt the experimental physicists of the
next generations, of many centuries onward - trying
to test whether this is indeed the "final" truth.
As a theorist, it is impossible to embrace, with the
great joy, the "final" truth without the deep
appreciation of its perfectness.

\medskip

\section{Concluding Remarks}

We are suggesting \cite{definition} that we live in the
quantum 4-dimensional Minkowski space-time with, {\it via
the gauge principle,} the force-fields gauge-group structure
$SU_c(3) \times SU_L(2) \times U(1)$ (at the fermi scale for
quarks), or $SU_L(2) \times U(1) \times SU_f(3)$ (at the
atomic scale for leptons), built-in from the very beginning.
The quark world, with the entries from the triplets of
the quark group $SU_Q(3)$, possesses the (123) gauge symmetry
and can be seen, at the fermi scale, by this overall
background. Meanwhile, the lepton world, with the entries
from the singlets of the quark group $SU_Q(3)$, possesses
another (123) gauge symmetry and can also be seen,
at the atomic scale, by this overall background.

We are using Einstein's relativity principle and the quantum
principle, established in the 20th Century, to describe the
{\it smallest} units of matter, such as electrons, neutrinos,
and quarks. It looks like an ultimate theory at its birth,
the so-called "Standard Model" \cite{definition}, which
we believe will eventually replace the thinkings during the
Newton's classic era (of the past four hundred years).

{\it Previously we underestimated the basic importance of the gauge 
principle and the fundamental importance of the matter group, or 
the quark group $SU_Q(3)$, in which the quark world are triplets and the 
lepton world are singlets. Once the gauge principle and the matter
group, both mathematical stuffs, are taken into account, the 
theory, or the Standard Model, is a consistent and perhaps 
complete theory. The concept of "generations" should not be there 
and thus it is simply a misnomer.}

In the quantum 4-dimensional Minkowski space-time, the
complex scalar field $\phi(x)$ is described renormalizably by
\begin{equation}
 {\cal L} = -(\partial_\mu \phi)^\dagger (\partial_\mu \phi)
 -M^2 \phi^\dagger \phi - \lambda (\phi^\dagger \phi)^2.
 \end{equation}
If $\lambda < 0$, the system collapses (i.e. unbounded from below).
If $\lambda > 0$, it is repulsive so that the system cannot build up
by itself. The interesting question is that $\lambda$ is dimensionless
- a pure number that characterizes the 4-dimensional Minkowski
space-time (maybe $\lambda ={1\over 8}$, but we need the proof),
{\it not} by the complex scalar fields.

The force fields are described by the gauge fields in either the
group $SU_c(3) \times SU_L(2) \times U(1)$ (for quarks), or the
group $SU_L(2) \times U(1) \times SU_f(3)$ (for leptons), {\it
via the gauge principle} - a mathematical principle (for the
completion of the description of motion). They need complex
scalar fields to generate the masses of the gauge bosons
via generalized Higgs mechanisms. The longitudinal components
are missing in the purely gauge-fields description such that
complex scalar fields are needed for the Higgs mechanisms.
Thus, in the quantum 4-dimensional Minkowski space-time with,
{\it via the gauge principle,} the force-fields gauge group
$SU_c(3) \times SU_L(2) \times U(1)$ (for quarks), or in the
gauge group $SU_L(2) \times U(1) \times SU_f(3)$ (for
leptons), the world is already set up in terms of the force
fields from the gauge fields and the complex scalar (Higgs)
fields via the specific Higgs mechanisms, and this
background world is called the "overall background".

It yields, and only yields, three Higgs fields $\Phi(1,2)$,
$\Phi(3,2)$, and $\Phi(3,1)$. The "related" Higgs fields, such as
$\Phi(1,2)$ and $\Phi(3,2)$ with $SU_L(2)$ doublets in common
(or, $\Phi(3,2)$ and $\Phi(3,1)$ with $SU_f(3)$ triplets in
common), can overcome the "self-repulsive" nature and become
useful and be able to live and to sustain in this world.

On the lepton world, we know that they couple to the
$SU_L(2) \times U(1)$ gauge sector (which makes them
visible). To interpret the ordering via three
"generations", we proposed the force-fields nature
of the $SU_f(3)$ gauge symmetry. Neutrino
oscillations provide a direct proof that "generations"
can switch among themselves. By introducing the
$SU_f(3)$ gauge symmetry to the original $SU_L(2) \times
U(1)$, the lepton world is free from the Landau ghost
and is asymptotically free. So, the lepton world is
perfectly well-behaved.

On the quark world, it is already a perfect world since
it couples to the $SU_c(3) \times SU_L(2) \times U(1)$ (i.e. the
standard (123)) gauge symmetry. It exhibits the "size" effect,
i.e., that the quark world exists only within a given volume
of fermi sizes;or, it also exhibits the temperature effect,
i.e., that it undergoes the phase transition (into something
else). The entry inputs are triplet members of the "quark"
group $SU_Q(3)$ - the entity with three "generations"
altogether serves one entry input.

{\it Thus, in this Standard Model, the old notion of
"generations" disappears altogether. The family
$SU_f(3)$ gauge symmetry and the presence of the
$SU_Q(3)$ symmetry both kills the notion of
"three generations".}

To sum up, we live in the quantum 4-dimensional Minkowski
space-time with, {\it via the gauge principle,} the
force-fields gauge group structure, $SU_c(3) \times
SU_L(2) \times U(1)$ (for quarks) or $SU_L(2) \times U(1)
\times SU_f(3)$ (for leptons), built-in from the very
beginning \cite{definition}. Apart from the $SSB$ "ignition"
term, the overall theory is both dimensionless and
massless \cite{Origin}. Neutrino oscillations are there.

\bigskip

\section*{Appendix: Sideline Remarks}

This world is, indeed,  very special. It is based on the quantum
4-dimensional Minkowski space-time with, {\it via the gauge
principle,} the force-fields gauge-group structure $SU_c(3)
\times SU_L(2) \times U(1)$, or $SU_L(2) \times U(1) \times
SU_f(3)$, built-in from the very beginning. We realize that
the complex scalar field is self-repulsive (i.e. does not exist)
if alone and that two "related" complex scalar fields could
interact attractively (and so exist) and they become the
very-much-wanted longitudinal components of the gauge fields.
The quark world, operating under the quark group $SU_Q(3)$
(not a gauge symmetry), would be accepted because of the
(123) gauge symmetry. Meanwhile, the lepton world could be
accepted in view of another (123) symmetry.

In this language, everything carries the group characteristic.
For example, $\nu_{e,L}$ belongs to the same multiplet as $e_L$,
so the same characteristic - thus, the case would be ruled out
that the electron is a Dirac particle while the neutrino is a
Majorana particle.

A few years ago \cite{Hwang3}, it was proposed that we
could work with two working rules: "Dirac similarity principle", based on
ninety years of experience, and "minimum Higgs hypothesis", from the last
fifty years of experience. Using these two working rules, the basic
model \cite{Hwang417} became rather unique in our choice - so, it is
so much easier to check it against the experiments. To move forward in
building up our knowledge, there are moments that we have to play
conservatively - including the moments of using these two working rules.

At this point, the two working rules seem to be rather trivial.

We have to say that the phenomenon of three generations used to be
one leading puzzle in particle physics; it is safe to add that neutrino
oscillations used to be another related puzzle. When we write
everything together, the need for introducing the Higgs $\Phi(3,2)$
becomes rather clear \cite{HwangYan}. As this world does not have
another massless gauge bosons (unless confined like gluons), it is
also clear that there is another pure family Higgs $\Phi(3,1)$ to
complete the story \cite{Family}. Is there any other possibility?
It seems that it is unique.

We have to admit that the natural consequence is the $SU_f(3)$ family gauge
theory - to put everything together, it leads to the (extended) Standard
Model \cite{Hwang417}, with a very natural generalized Higgs mechanism
for the three Higgs. It is in accord with the "minimum Higgs hypothesis".
These complex scalar fields are meant together and interact together, and
nothing more.

It is also of importance to point out that there is no experimental
evidence to the assertion that neutrinos are point-like Dirac
particles - "Dirac similarity principle" should be in experimental
checks. Of course, they are in the same multiplet, like in
\cite{HwangYan}, so that they both are Dirac particles of the
same nature. So, the two working principles are justified by
our experiences for ninety or forty years. Surprising enough,
it seems to work this way, at least so far.

Thus, we may declare that our world is the quantum 4-dimensional
Minkowski space-time with, {\it via the gauge principle,} the
force-fields gauge-group structure $SU_c(3) \times SU_L(2)
\times U(1)$, or $SU_L(2) \times U(1) \times SU_f(3)$, built-in
at the very beginning; in our world, the various Higgs "exist"
such that all gauge bosons, except the photon, are either
confined or massive - this provides the overall background
to support the "triple" quark world as well as to support the
lepton world. This is the origin of "point-like particles", or
of "smallest particles", or of "fields".

To close up our discussions, there remain a few philosophical questions
in our minds: First, why are the complex scalar fields so special in
the 4-dimensional Minkowski space-time? The 4-dimensional
Minkowski space-time (not the complex scalar fields themselves)
would be the reason to determine the value of
$\lambda$ ($={1\over 8}?$). In the other dimensions
(different from four), the existence of the self-repulsive
interaction $\lambda (\phi^\dagger\phi)^2$ would not have
the same status. Second, why do we need the ignition
channel, which turns out to be in the purely family channel
($\mu_2^2< 0$)? Third, the dimensional regularization, or others,
might not give whole story regarding the (leading) ultraviolet
divergences. In the beginning of this 21th century, we might
have this ghost (infinite or divergence) story
to resurface again \cite{fine-tune, definition}. But this
might be destined to be so, as our knowledge accumulates in
the process.

We may add that, under two working hypotheses or under our Standard
Model (in the quantum 4-dimensional Minkowski space-time with the
$SU_c(3) \times SU_L(2) \times U(1)$, or $ SU_L(2) \times U(1)
\times SU_f(3)$, gauge-group structure built in from the very
beginning), we should be able to close the Universe; that is,
all the dark-matter particles and all the ordinary-matter
particles are accounted for. Our Standard Model provides a
description of the entire matter world - i.e., the
25\% dark-matter world and the 5\% ordinary-matter world.

In our language, the {\it a priori} world (i.e., the
overall background) is the quantum 4-dimensional Minkowski
space-time with some force-fields gauge-group structure
built-in from the very beginning. The triple quark world
and the lepton world are something added on at a later
stage (at the different scales), and in fact they may be
the added-on at a different stage and at the fermi or
atomic scales - that makes the study of leptons a subject
by itself \cite{collider}.

We would be curious about how the dark-matter world looks like, though
it is difficult to verify experimentally. The first question would be: The
dark-matter world, 25 \% of the current Universe (in comparison, only 5 \%
in the ordinary matter), would clusterize to form the dark-matter
galaxies? The dark-matter galaxies would then play the hosts of
(visible) ordinary-matter galaxies, like our own galaxy, the Milky
Way. Note that a dark-matter galaxy is by our definition a galaxy
that does not possess any ordinary strong and electromagnetic
interactions (with our visible ordinary-matter world). These
fundamental question deserves some in-depth thoughts, for the
evolution of our Universe.

The situation may be relatively simple, if we look closely at
our Standard Model \cite{definition}. All the dark-matter candidates,
except neutrinos and antineutrinos, will decay away with a lifetime
shorter than a fraction of a sub-second \cite{HwangP1}. Since neutrinos
have masses, heaviest of $0.058\,eV$, the cosmic background neutrinos
(CB$\nu$'s) would cluster around the visual ordinary-matter centers.
We believe that the CB$\nu$'s account for all of the $25\%$ dark matter.

Of course, we should remind ourselves that, in our ordinary-matter world,
those quarks can aggregate in no time, to hadrons, including nuclei, and
the electrons serve to neutralize the charges also
in no time. Then atoms, molecules, complex molecules, and so on. These serve as
the seeds for the clusters, and then stars, and then galaxies, maybe in a time span
of $1\, Gyr$ (i.e., the age of our young Universe). The aggregation caused by
strong and electromagnetic forces is fast enough to help giving rise to galaxies
in a time span of $1\, Gyr$. {\it We believe that all these took place in the
invisible environments of neutrino halos.}

In other words, the neutrino halos, the $25\%$ dark matter, play
the invisible roles of our Universe. They are there, very light
but definitely massive, could not have the very heavy invisible
centers, and use the visible heavy centers (such as planets,
stars, etc.) as their centers. This would modify the Newton's
gravitational law, a macroscopic law, with the invisible
parts. The invisible stories may sound scary, but after
certain clarifications everything goes its own way except
that there is some invisible partner(s).

\end{document}